\documentclass[aps, 10pt, preprintnumbers,prd, amsmath,amssymb,notitlepage]{revtex4-1} %, ,10pt
\usepackage{graphicx}% Include figure files latexsym,array,enumerate,letter,numbers,twocolumn
\usepackage{bm}% bold math
\usepackage{amssymb}
\usepackage{epsfig}
\usepackage{color}
\usepackage{mathtools}
\usepackage{cases}
\usepackage{booktabs}
\usepackage{amsmath}
\usepackage{subcaption}
 \usepackage{url}
\usepackage[
colorlinks=true,
filecolor=black,
anchorcolor=blue,
linkcolor=blue,
citecolor=cyan, %red,
urlcolor=cyan,
linktocpage=true,
plainpages=false,
breaklinks=true,
pdfstartview=FitH
]{hyperref}

\newcommand{\ck}[1]{\textcolor{black}{#1}}
\newcommand{\ckk}[1]{\textcolor{black}{#1}}
\newcommand{\vev}[1]{\left<{#1}\right>}

\newcommand{\w}{\omega}
\newcommand{\wg}{\omega_g}

\newcommand{\be}{\begin{equation}}
\newcommand{\ee}{\end{equation}}
\newcommand{\bea}{\begin{equation}\begin{aligned}}
\newcommand{\eea}{\end{aligned}\end{equation}}

\newcommand{\PBH}{{\scriptscriptstyle \textrm{PBH}}}

\begin{document}
\title{ Novel high-frequency gravitational waves detection with split cavity}
\author{Chu-Tian Gao$^{1}$}
%\email{your@email, \& please check affliations}
\author{Yu Gao$^{2}$}
%\email{gaoyu@ihep.ac.cn}
\author{Yiming Liu$^{1}$}
%\email{your@email,\& please check affliations}
\author{Sichun Sun$^{1}$}
%\email{sichunssun@bit.edu.cn}
%\author{Yun-Long Zhang$^{3,4,5}$}
%\email{zhangyunlong@nao.cas.cn}

\affiliation{$^1$School of Physics, Beijing Institute of Technology, Beijing, 100081, China}
\affiliation{$^{2}$Key Laboratory of Particle Astrophysics, Institute of High Energy Physics,
Chinese Academy of Sciences, Beijing 100049, China}
%\affiliation{$^3$National Astronomy Observatories, Chinese Academy of Science, Beijing, 100101, China}
%\affiliation{$^4$School of Fundamental Physics and Mathematical Sciences, Hangzhou Institute for Advanced Study, University of Chinese Academy of Sciences, Hangzhou 310024, China}
%\affiliation{$^5$International Center for Theoretical Physics Asia-Pacific, Beijing/Hangzhou, China}

\begin{abstract}
Gravitational waves can generate electromagnetic effects inside a strong electric or magnetic field within the Standard Model and general relativity. Here we propose using a quarterly split cavity and LC(inductor and capacitor)-resonance circuit to detect a high-frequency gravitational wave from 0.1 MHz to GHz. We perform a full 3D simulation of the cavity's signal for sensitivity estimate. Our sensitivity depends on the coherence time scale of the high-frequency gravitational wave sources and the volume size of the split cavity.
We discuss the resonant measurement schemes for narrow-band gravitational wave sources and also a non-resonance scheme for broadband signals. For a meter-sized split cavity under a 14 Tesla magnetic field, the LC resonance enhanced sensitivity to the gravitational wave strain is expected to reach $h\sim 10^{-20}$ around $10$ MHz.
\end{abstract}

\maketitle

\section{introduction}

LIGO and Virgo, the ground base interferometers have directly observed the gravitational waves(GWs) \cite{LIGOScientific:2016aoc} , which started the era of GW astronomy. Along with electromagnetic waves from another long-range interaction, now we can observe the universe from both spectra. The central focus of the current GW experiment is from Hz to kHz with ground-based experiments\cite{Hild:2010id, Punturo:2010zz,LIGOScientific:2016wof}.  The other gravitational waves proposal, from high to ultra-low frequency, includes space-based\cite{LISA:2017pwj,Yagi:2011wg}, moon-based\cite{LGWA:2020mma,vanHeijningen:2023esw}, laser/atom interferometers\cite{Badurina:2019hst,Abe:2021ksx,AEDGE:2019nxb}, pulsar timing arrays\cite{NANOGrav:2020bcs,Janssen:2014dka}, and CMB observations\cite{Namikawa:2019tax,CMB-S4:2020lpa}, etc.

If we go into the other end of the gravitational wave spectra, the higher frequency, various interesting proposals, and detectors have already been put forth, including superconducting rings\cite{Anandan:1982is},  microwave and optical cavities\cite{Mensky:2009zz, Caves:1979kq, Pegoraro:1978gv,Pegoraro:1977uv, Reece:1984gv, Reece:1982sc, Ballantini:2005am,Bernard:2002ci,Bernard:2001kp, Ballantini:2003nt, Cruise:2000za, Cruise:2005uq, Cruise:2006zt}, interferometers.\cite{Ackley:2020atn, Bailes:2019oma, Akutsu:2008qv, Holometer:2016qoh, Nishizawa:2007tn, sym14102165}, optically levitated sensors\cite{Aggarwal:2020umq}, mechanical resonators\cite{Goryachev:2014nna,Goryachev:2014yra, Aguiar:2010kn, Gottardi:2007zn}, and detectors based on the inverse-Gertsenshtein effect\cite{Gertsenshtein1962,Braginskii:1973vm} and the magnon modes\cite{Ito:2019wcb,Ito:2022rxn}( e.g. see \cite{Aggarwal:2020olq} for a comprehensive review), and some very recent novel proposals \cite{Bringmann:2023gba,Domcke:2020yzq,Barrau:2023kuv,Vadakkumbatt:2021fnw,Howl:2021giy,Goryachev:2021zzn}. Those experiments are designed to be tabletop or room-size, to match the smaller wavelength. However, there are still orders of magnitudes in gravitational wave amplitude strains for experiments to meet theory prediction.

Recent progress has been made that existing axion search experiments can already cast the limits on the high-frequency gravitational waves sensitivities \cite{Domcke:2022rgu,Berlin:2021txa}. In this work, we propose a new way to detect high-frequency gravitational waves from kHz to GHz, a design with a split cavity as a capacitor and readout LC circuit, different yet resembling those axion detection schemes, such as cavity-based ADMX\cite{ADMX:2021nhd}, non-cavity ABRACADABRA\cite{ABRACADABRA:2018rtf,Salemi:2019xgl}, SHAFT\cite{Gramolin:2020ict} and DM-Radio\cite{Brouwer:2022bwo,Silva-Feaver:2016qhh}. The quarterly split cavity roughly matches the quadruple shape of the gravitational wave oscillations and does not need to resemble the geometric rigidity of a closed cavity. Its frequency response does not need to develop a sharp peak at the resonant frequency, thus it can also be used in a broad-band measurement. The high-frequency gravitational wave detection proposal here employs a similar readout and capacitor-based design, with the axion detection RELEAP\cite{Duan:2022nuy}.  For most of the proposed gravitational wave sources, signals do not have a high coherent factor, which is the crucial difference from the axion oscillating background. Here we discuss different signal processing schemes and show both the exclusion limits for the LC-resonant scheme, targeting coherent narrowband sources, and the non-resonant broadband scheme for transient signals.

We organize the rest of the paper as follows: in section II, we show the theoretical calculation of the gravitational wave-induced electromagnetic current, especially the result for incoming gravitational waves in arbitrary angles; in section III, we discuss the electromagnetic solution with the split cavity and the numerical simulation with COMSOL; in section IV, we show the sensitivity reach of our proposal, based on the numerical simulation, and comparing them with the other limits with broadband axion detectors.

\section{Gravitational Wave Electrodynamics}
Throughout this paper, we use Heaviside units with $\hbar= c = 1$, $\eta_{\mu\nu} = \mathrm{diag}(- +++)$, work to leading order in $|h_{\mu\nu}|$, and raise indices with $\eta^{\mu\nu}$.

%\subsection{theoretical set-ups}
	We start with the Lagrangian for the Inverse Gertsenshtein effect, which is present in the Standard Model~\cite{Ejlli:2019bqj}:
$$	
	\begin{aligned}
		S=\int d^4 x \sqrt{-g}\left(-\frac{1}{4} g^{\mu \alpha} g^{\nu \beta} F_{\mu \nu} F_{\alpha \beta}\right) 
		\end{aligned}
$$	
Then we can expand the metric to the leading order in $h$, as below
$$		
	\begin{aligned}
		S \supset-\frac{1}{2} \int d^4 x j_{\mathrm{eff}}^\mu A_\mu
		\end{aligned}
$$	
where
		$$
	j_{\mathrm{eff}}^\mu \equiv \partial_\nu\left(\frac{1}{2} h F^{\mu \nu}+h_\alpha^\nu F^{\alpha \mu}-h_\alpha^\mu F^{\alpha \nu}\right)
	$$
	
		Here we employ the result from Ref.\cite{Berlin:2021txa} with a most positive metric notation. For convenience, we can write out $h_{\mu \nu}$ in the TT frame,
	$$
	\begin{aligned}
		& h_{00}=-\omega_g^2 h_{a b}^{\mathrm{TT}} x^a x^b\left[\frac{i}{-\omega_g z}+\frac{1-e^{-i \omega_g z}}{\left(\omega_g z\right)^2}\right] \\
		& h_{i j}=\omega_g^2\left[\left(\delta_{i z} h_{j a}^{\mathrm{TT}}+\delta_{j z} h_{i a}^{\mathrm{TT}}\right) z x^a-h_{i j}^{\mathrm{TT}} z^2-\delta_{i z} \delta_{j z} h_{a b}^{\mathrm{TT}} x^a x^b\right]\left[-\frac{1+e^{-i \omega_g z}}{\left(\omega_g z\right)^2}-2 i \frac{1-e^{-i \omega_g z}}{\left(\omega_g z\right)^3}\right] \\
		& h_{0 i}=-\omega_g^2\left(h_{i a}^{\mathrm{TT}} z x^a-\delta_{i z} h_{a b}^{\mathrm{TT}} x^a x^b\right)\left[\frac{-i}{2 \omega_g z}-\frac{e^{-i \omega_g z}}{\left(\omega_g z\right)^2}-i \frac{1-e^{-i \omega_g z}}{\left(\omega_g z\right)^3}\right],
	\end{aligned}
	$$

	The components of $h_{i j}^{\mathrm{TT}}$  can then take the following explicit forms
	$$
	\begin{aligned}
		&\left.h_{\rho \rho}^{\mathrm{TT}}\right|_{\mathbf{r}=0}=-e^{i \omega_{g} t}\left(-h^{+}\left(\sin ^2\left(\phi-\phi_h\right)-\cos ^2\left(\phi-\phi_h\right) \cos ^2 \theta_h\right)+2 h^{\times} \cos \theta_h \cos \left(\phi-\phi_h\right) \sin \left(\phi-\phi_h\right)\right), \\
		&\left.h_{\rho \phi}^{\mathrm{TT}}\right|_{\mathbf{r}=0}=-e^{i \omega_{g} t}\left(-h^{+}\left(1+\cos ^2 \theta_h\right) \sin \left(\phi-\phi_h\right) \cos \left(\phi-\phi_h\right)+h^{\times} \cos \left(2\left(\phi-\phi_h\right)\right) \cos \theta_h\right), \\
		&\left.h_{\rho z}^{\mathrm{TT}}\right|_{\mathbf{r}=0}=e^{i \omega_{g} t}\left(h^{+} \cos \theta_h \sin \theta_h \cos \left(\phi-\phi_h\right)+h^{\times} \sin \theta_h \sin \left(\phi-\phi_h\right)\right), \\
		&\left.h_{\phi z}^{\mathrm{TT}}\right|_{\mathbf{r}=0}= -e^{i \omega_{g} t}\left(h^{+} \cos \theta_h \sin \theta_h \sin \left(\phi-\phi_h\right)-h^{\times} \sin \theta_h \cos \left(\phi-\phi_h\right)\right), \\
		&\left.h_{z z}^{\mathrm{TT}}\right|_{\mathbf{r}=0}= -e^{i \omega_{g} t} h^{+} \sin ^2 \theta_h .
	\end{aligned}
	$$
	
	We here follow the notation in Ref.\cite{Domcke:2022rgu,Berlin:2021txa}, and further explicitly calculate out the gravitational wave effective current with incoming gravitational waves in arbitrary angles.
Let us take the background magnetic field to be static, spatially uniform, and pointing along the $+\hat{z}$ direction in the TT frame, $\mathbf{B}_{0}=B_{0}\hat{z}$,  $j^{\mu}_{\mathrm{eff}}$ can be determined by direct calculation,
\begin{equation}
	\begin{aligned}
	j^{0}_{\mathrm{eff}}&=\frac{B_{0}(-1+e^{\kappa})w_{g}^{2}(\mathrm{k_{1}b_{2}-b_{1}k_{2}})}{-\kappa}\\
	j_{\mathrm{eff}}^{\rho}&= \frac{1}{2 \kappa^3} B_{0} w_{g}^2(2+\kappa)\left(-2-2 e^\kappa(-1+\kappa)+\kappa^2\right) \mathrm{b_{2}}+\frac{1}{\kappa^2} B_{0} w_{g}^2\left(1+e^\kappa(-1+\kappa)\right) \rho h^{\mathrm{TT}}_{\rho\phi}|_{\mathbf{r}=0} \\
	&-\frac{1}{2 \kappa^4} i B_{0} w_{g}^3\left(-6+e^\kappa(6-4 \kappa)-2 \kappa+\kappa^2+\kappa^3\right)(\mathbf{b} \cdot \mathbf{r})\mathrm{k_{2}}\\
	 j_{\mathrm{eff}}^{\phi}&=\frac{B_{0} w_{g}^2\left(1+e^\kappa(-1+\kappa)\right) \rho}{\kappa^2} h^{\mathrm{TT}}_{\phi\phi}|_{\mathbf{r}=0}+\frac{i B_{0} w_{g}^3\left(-6+e^\kappa(6-4 \kappa)-2 \kappa+\kappa^2+\kappa^3\right)(\mathbf{b} \cdot \mathbf{r}) \mathrm{k_{1}}}{2 \kappa^4} \\
	& -\frac{B_{0} w_{g}^2(2+\kappa)\left(-2-2 e^\kappa(-1+\kappa)+\kappa^2\right)\mathrm{b_{1}}}{2 \kappa^3}\\
		j_{\mathrm{eff}}^{z}&=\frac{B_{0} w_{g}^2\left(1+e^\kappa(-1+\kappa)\right) \rho h^{\mathrm{TT}}_{\phi z}|_{\mathbf{r}=0}}{\kappa^2}
	\end{aligned}
\end{equation}

Where $\mathbf{b}=(b_{1}, b_{2}, b_{3}), b_{j}=r_{i}h^{\mathrm{TT}}_{ij}|_{\mathbf{r}=0}, \mathbf{k}=(k_{1}, k_{2}, k_{3})=(\sin\theta_{h}\cos(\phi-\phi_{h}), -\sin\theta\sin(\phi-\phi_{h}), \cos\theta_{h}), \mathbf{r}=(\rho,\phi, z), \kappa=-i \mathbf{k.r}$.

\begin{figure}
    \centering
    \begin{subfigure}{.5\textwidth}
        \centering
    \includegraphics[scale=0.3]{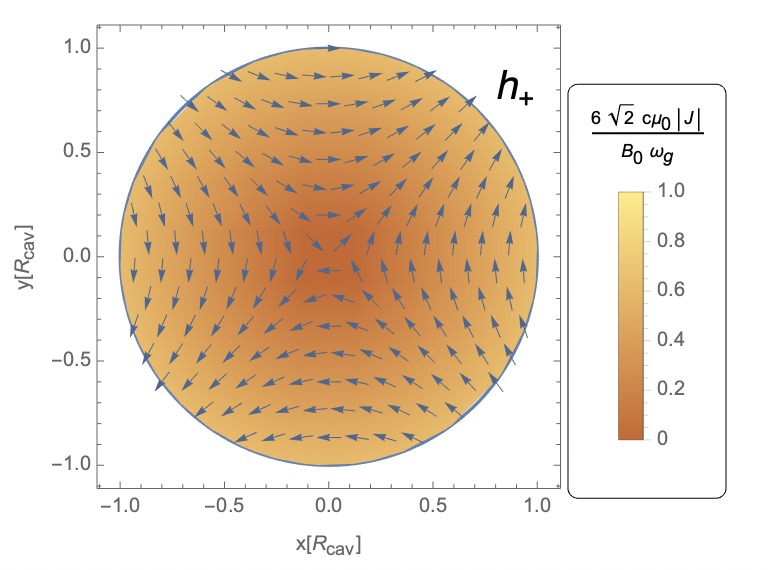}
    \end{subfigure}%
    \begin{subfigure}{.5\textwidth}
        \centering
      \includegraphics[scale=0.3]{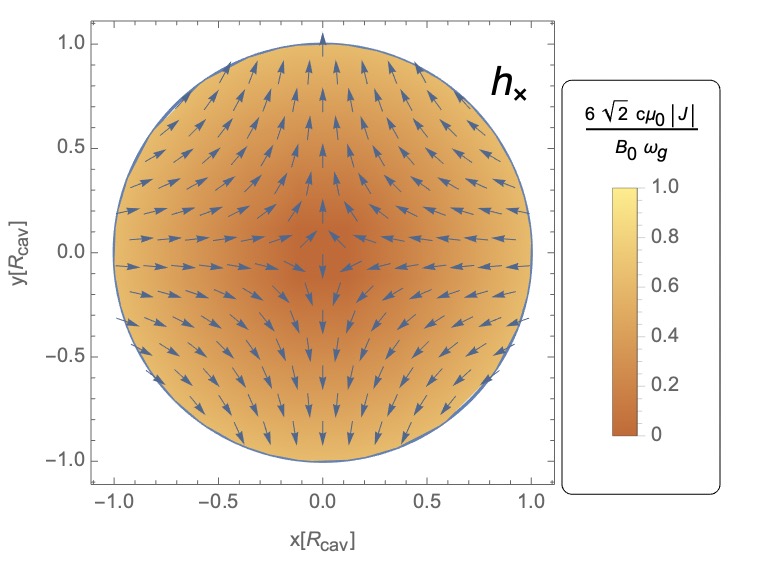}
    \end{subfigure}
    \caption{We plot the $j_\text{eff}$ with only cross/plus mode at z=0 in x-y plane in Eq.\ref{eq:jeff1}. The arrows denote the direction of $j^\mu$ and the different color scales represent the magnitude.
    %TE$_{211}$ is the lowest mode that satisfies the spin-2 rotation symmetry of the gravitational wave perturbation.
    }
    \label{fig:jeff}
\end{figure}

	If we take the incoming gravitational waves direction as the z-direction, aligned with the magnetic field, and assume cylindrical symmetry with cylindrical coordinates, the GW effective current can be simplified as:
	%For this specific example, the direction of the $\mathrm{GW}$ and the applied magnetic field preserve the cylindrical symmetry of the magnetic field. As a result, the form of $j_{\text {eff }}^\mu$ vastly simplifies, This is made manifest by rewriting $j_{\text {eff }}^\mu$ in cylindrical coordinates, i.e., $j_{\mathrm{eff}}^\mu=\left(\rho_{\mathrm{eff}}, j_{\mathrm{eff}}^r, j_{\mathrm{eff}}^\phi, j_{\mathrm{eff}}^z\right)$, such that
	\begin{align}
	j_{\mathrm{eff}}^\mu=-\frac{B_0 \omega_g^2 r}{6 \sqrt{2}} e^{i \omega_g t}\left(0, i e^{-2 i \phi} h_{+2}-i e^{2 i \phi} h_{-2}, e^{-2 i \phi} h_{+2}+e^{2 i \phi} h_{-2}, 0\right) \times f\left(\omega_g z\right), \label{eq:jeff1}
	\end{align}
	where $f(x) \equiv-3-6 i x^{-1}-12 e^{-i x} x^{-2}-12 i\left(1-e^{-i x}\right) x^{-3}$ is a dimensionless function with $\lim _{x \rightarrow 0} f(x)=1$ and $\lim _{x \rightarrow \infty} f(x)=-3$. GW helicity components are defined as $h_{ \pm 2} \equiv\left(h_{+} \pm i h_{\times}\right) / \sqrt{2}$. %which transform under a rotation by $\Delta \phi$ about the cavity/GW axis as $h_{ \pm 2} \rightarrow e^{ \pm 2 i \Delta \phi} h_{ \pm 2}$.
In the SI unit, we can replace $B_0$ with $B_0 /\left(\mu_0 c^2\right)$ above in $j_{\text {eff}}^\mu$. We illustrate $j_\text{eff}$ in the x-y plane in Fig.\ref{fig:jeff}.

\section{Solutions with split cavity}

\subsection{Cylindrical modes inside a solenoid}\label{app:solenoidal_modes}

A solenoid is the common form of an experimental strong $B$-field. We would first find out the induced electromagnetic modes under the symmetries of a solenoid, and design the readout apparatus accordingly. 

The eigenmode solutions inside a closed cylinder of radius $R_{\rm cav}$ and finite length $L_{\rm cav}$, under ideal conductor boundary conditions, are classified into resonant transverse magnetic (TM) and transverse electric (TE) modes~\cite{Hill}. The TM modes are:
\begin{eqnarray}
	E_r^\pm &= &- \left[{A_+ \sin m\phi \atop A_-\cos m\phi}\right]\frac{k_z}{\w_{mnp}^2-k_z^2}\sin(k_zz)J'_{m}(r\gamma_{mn})\gamma_{mn},\label{eq:TM_r}\\
	E_\phi^\pm&=&-\left[{A_+ \cos m\phi \atop -A_-\sin m\phi}\right]\frac{ k_z}{\w_{mnp}^2-k_z^2}\sin(k_zz)\frac{m}{r}J_{m}(r\gamma_{mn}),\label{eq:TM_phi}\\
	E_z^\pm&=& \left[{A_+ \sin m\phi \atop A_- \cos m\phi}\right]J_{m}(\gamma_{mn} r)\cos(k_zz),\label{eq:TM_z}
\end{eqnarray}
where $\omega_{mnp}$ is the frequency at the resonant mode. The $\pm$ at the $E$-fields correspond to the upper/lower component in the brackets, $A_\pm$ are constants that are determined by normalization condition.  We have $k_z=\pi p/L_{\rm cav}$, $\gamma_{mn}=x_{mn}/R_{\rm cav}$, $x_{mn}$ is the $n$th zero of 1st Bessel function $J_m$. The mode numbers $m$ and $p$ are non-negative integers while $n$ is a positive integer.
The TE$_{mnp}$ modes are:
\begin{eqnarray}
    E_r^\pm&=&\left[{A_+ \cos m\phi \atop -A_- \sin m\phi}\right]\sin \left(k_zz\right)\frac{i\w_{mnp} }{\w_{mnp}^2-k_z^2} \frac{m}{r} J_{m}(r\gamma_{mn}),\label{eq:TE_r}\\
    E_\phi^\pm&=&-\left[{A_+ \sin m\phi \atop A_- \cos m\phi}\right]\frac{i\w_{mnp} }{\w_{mnp}^2-k_z^2}\sin \left(k_zz\right) J_{m}'(\gamma_{mn} r)\gamma_{mn},\label{eq:TE_phi}\\
    E_z^\pm&=&0,\label{eq:TE_z}
\end{eqnarray}
where $\gamma_{mn}=x_{mn}'/R_{\rm cav}$, and $x_{mn}'$ is the $n$th zero of $J'_m$. Those modes are orthogonal to each other as defined in the mode bases. These modes are used later to demonstrate the spin-2 nature of the effective current.
%In the main text we denote the mode functions by $\E_{mnp\pm}^{\rm TM}$ or $\E_{mnp\pm}^{\rm TE}$. Often we drop the superscript TM and TE because it is clear from the context what we mean. TM and TE modes are orthogonal to each other in the sense that $\int_{\Vcav} dV\, \E^{\rm TM}\cdot \E^{\rm TE}=0$, and likewise the $+$ and $-$ modes are degenerate and orthogonal for the same $m \neq 0$, $n$, and $p$. Note that an equivalent formalism is to define the modes in a basis of eigenfunctions of the azimuthal rotation generator $\partial_\phi$, which amounts to taking linear combinations of $\E_+$ and $\E_-$; demonstrating the spin-2 nature of the effective current.

\subsection{A quarterly split cavity}

In principle, any induced signal EM field can be expressed as the linear combination of the modes in Eq.~\ref{eq:TM_r}-\ref{eq:TE_z}. If we assume a gravitational plane wave that propagates along the cylinder's $\hat z$ axis, the effective current $j_{\rm eff}$ lies in the transverse $\hat{x}-\hat{y}$ directions. %and \ckk{incite mostly TE modes inside the cavity.}
The spin-2 nature of the gravitational mode is reflected by the $e^{\pm i 2\phi}$ rational symmetry in Eq.~\ref{eq:jeff1}.
%which sets the base mode in $\phi$-rotation around the axis, and any higher TE modes must come at multiples of this base mode in the $\phi$ direction.

Taking the lowest ${\bf E}$ mode numbers, the base mode that the gravitational waves along the z direction excite is $TE_{211\pm}$ and its spatial components are given as
\begin{eqnarray}
    E_r^\pm&=&\left[{A_+ \cos 2\phi \atop -A_- \sin 2\phi}\right]\sin \left(\frac{\pi z }{L_{\rm cav}}\right)\frac{i\w_{211} }{\w_{211}^2-(\frac{2\pi }{L_{\rm cav}})^2} \frac{2}{r} J_{2}(r\gamma_{21}),\label{eq:TE_r211}\\
    E_\phi^\pm&=&-\left[{A_+ \sin 2\phi \atop A_- \cos 2\phi}\right]\frac{i\w_{211} }{\w_{211}^2-(\frac{2\pi }{L_{\rm cav}})^2}\sin \left(\frac{\pi z}{L_{\rm cav}}\right) J_{2}'(\gamma_{21} r)\gamma_{21},\label{eq:TE_phi211}\\
    E_z^\pm&=&0,\label{eq:TE_z211}
\end{eqnarray}
and the corresponding resonant frequency is
\begin{eqnarray}
\w_{211}=\sqrt{k_z^2+\gamma_{mn}^2}=\sqrt{(\frac{\pi }{L_{\rm cav}})^2+(\frac{x^\prime_{21}}{R_{\rm cav}})^2}.
\end{eqnarray}
The vertical components $E_\phi, E_z$ vanish at the closed boundary, $r=R_{cav}$. Only $E_r$ is non-zero and it is perpendicular to the cylindrical inner surface.

\begin{figure}
    \centering
    \begin{subfigure}{.5\textwidth}
        \centering
    \includegraphics[scale=1.2]{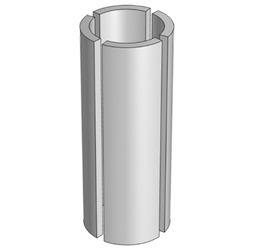}
    \end{subfigure}%
    \begin{subfigure}{.5\textwidth}
        \centering
      \includegraphics[scale=0.2]{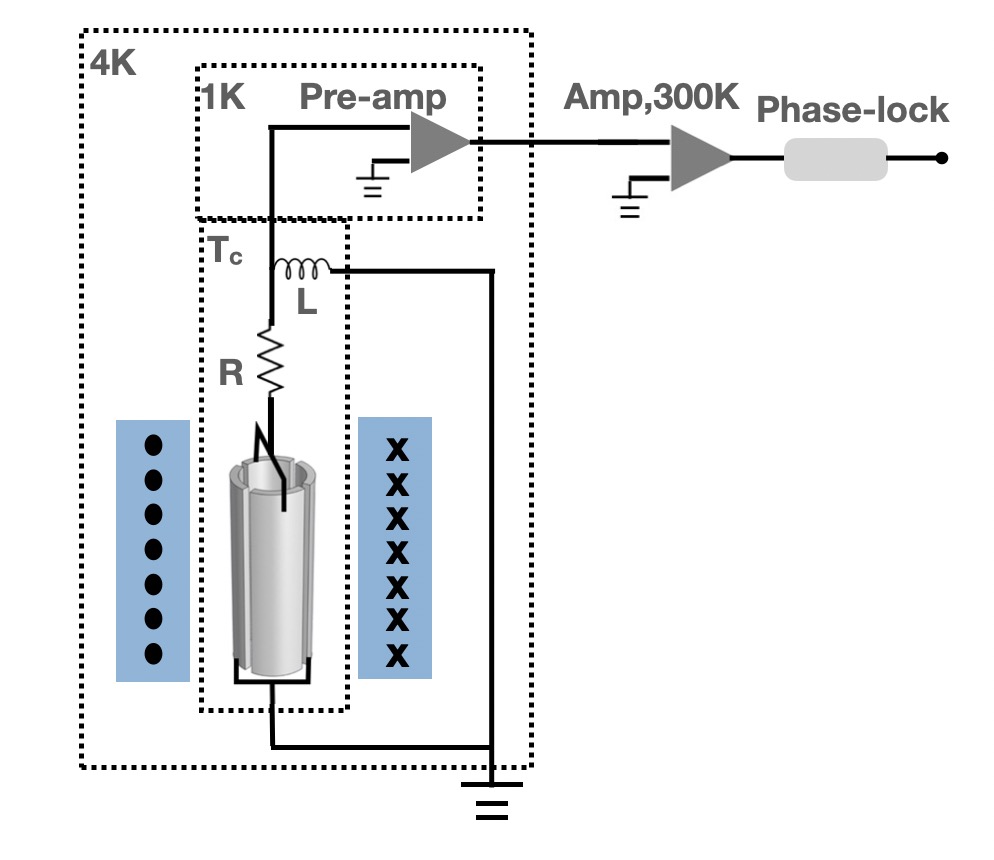}
    \end{subfigure}
    \caption{Split cavity: the top, bottom, and quarterly split periphery of the cylinder cavity are separated by insulation to form induction signal currents through the connection wiring. The slit angle is $3^{\circ}$ in this case. The connecting wires and amplification LC circuit are also shown.
    %TE$_{211}$ is the lowest mode that satisfies the spin-2 rotation symmetry of the gravitational wave perturbation.
    }
    \label{fig:quarterly}
\end{figure}

For a given cylinder size, the optimal situation is when the signal frequency $\omega$ matches one of the TE modes'. Frequency mismatch between the signal and cavity's eigenmodes leads to non-zero projection into higher TE modes, which may not always sum up constructively. As a consequence, efficiency loss will occur in both high-frequency and low-frequency directions:  $\omega \ll \omega_{211}$ is expected to suffer major form factor suppression, and $\omega \gg {\rm min}(R_{\rm cav}^{-1}, L_{\rm cav}^{-1})$ will hit the decoherence limit due to the signal photon conversion in a spatial region larger than its wavelength. The maximal conversion efficiency typically occurs near the base mode $\omega_{211}$ for the given dimensions of a cavity. Therefore, in a narrow-band setup, it favors applying frequency filtering in a neighborhood around $\omega\sim \omega_{211}$.

The electric field distribution will redistribute the electrons on the cylinder's conductor surface, and TE$_{211}$ mode will feature in a quadruple-like distribution in the cross-section of the cylinder. We split up the cylinder in a quarterly manner so that the electron redistribution between two adjacent plates has to flow through external wiring, thus leading to a measurable current signal, as illustrated in Fig.~\ref{fig:quarterly}. The resulting electric response does not depend on the particular narrow slit angle drastically. Note the quarterly pattern will mostly pick up the quadruple component, the plus or cross mode that aligns with the slits' $\phi$-separation, and loses the orthogonal component that leans at a 45$^\circ$ angle.
%hence a factor of \ckk{$1/2$?} loss will apply.

\begin{figure}[t]
    \centering
    \includegraphics[scale=0.35]{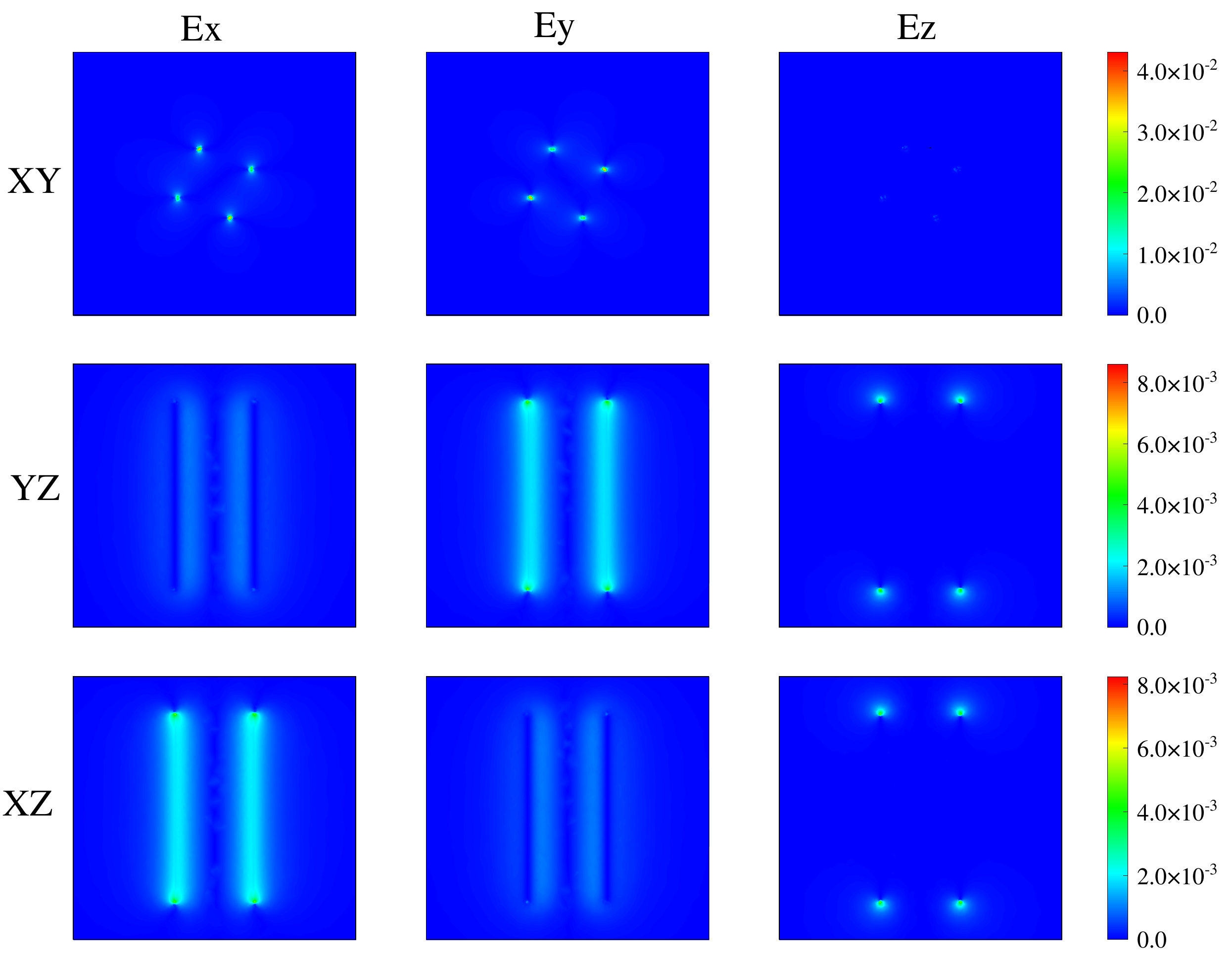}
    \caption{EM response simulation of the incoming gravitational waves $h=10^{-10}$ in the z-direction with a quarterly split cylinder $R=1$cm and $ L=5$cm.  The unit is $V/m$.}
    \label{fig:EM_sim}
\end{figure}

Splitting up the plates will cause significant deformation on the TE$_{211}$ mode. Even when the slits are narrow, their impact on the eigenmodes is noticeable, mostly through the aperture enhancement effect as in Ref.\cite{Harrington}. We performed an EM numerical simulation of the split cylinder with COMSOL package~\cite{COMSOL} to obtain the electric field distribution within the cavity and then compute the corresponding electric charge distribution on the quarter-plates at different frequencies. With an LC enhancement, we do not require the cavity itself to provide narrow-band filtering. Thus in principle, the cavity does not need to resemble the geometric rigidity of a closed cavity. In other words, a split/open cavity's frequency response does not need to develop a sharp peak at $\omega_{211}$ and it can also be used in a broad-band measurement, which we will discuss later.
%and found good agreement with a closed cylinder's TE$_{211}$ mode. 

\ckk{Besides its simple geometry, the quarterly split electrodes conveniently enclose the volume inside the solenoid, this is convenient for good signal efficiency as the maximal amount of the signal's conversion power is proportional to the magnetized volume. Since our split cavity is open at the top and the bottom, the aperture effect will cause some field-line leakage through these openings, leading to a sub-unity geometry form factor if compared to a completely closed resonant cavity. A relatively large length-to-aperture ratio will approach the solenoid symmetry that helps to reduce the loss at open ends, which can be evaluated via EM simulations.}

The EM simulation results for the plates' charge accumulation and effective capacitance are shown in Table~\ref{tab:charge_on_plate}.  The set-up with the split cavity and LC circuit amplification readout is shown in Fig.\ref{fig:quarterly}, and the full numerical EM simulation is shown in Fig.\ref{fig:EM_sim}. We refer to Ref. \cite{Duan:2022nuy} for more details of the cryogenic amplification readout system.
%\ck{\bf [More details are needed: (1) what is the EM sim setup; (2) What is the R value (capacitance relies on dimension, not just the R:d ratio) (3) the corresponding 211 frequency with the chosen plate size]}

\begin{table}[h]
\begin{tabular}{c|c|c||c|c|c}
\hline
\multicolumn{3}{c|}{R:L = 1cm:5cm} & \multicolumn{3}{|c}{R:L = 1m:1m~} \\
\hline
$f$ (Hz)& $C$ (pF) & $q$ (C) &$f$ (Hz)& $C$ (pF) & $q$ (C) \\
\hline
$10^7$ &  1.53 & $2.39\times10^{-19}$    &$10^5$ & 31.05 & $2.04\times10^{-16}$\\
$10^8$ &1.53  & $2.39\times10^{-18}$    &$10^6$  & 31.05  & $2.04\times10^{-15}$\\
$10^9$ &  1.52  & $2.29\times10^{-17}$    &$10^7$  & 31.05 & $2.03\times10^{-14}$\\
\hline
\end{tabular}
\caption{Selected values of the cavity's effective capacitance $C$ and the accumulated charge $q$  from EM simulations. The GW is assumed to propagate along the cavity axis with an amplitude $h_0= 10^{-10}$ . The left and right columns show two experimental scales: a multi-center-sized cavity with $f_{211}= 1.5\times10^{10}$ Hz, and a meter-sized cavity with $f_{211}=2.1\times 10^8$ Hz. Notice here our working frequencies are away from the closed cavity's resonance frequencies.
%\cs{for $R$=1m, $d$=1m and $R$=1cm, $d$=5cm}, respectively. The frequencies are chosen according to $ \omega_a = \frac{2\pi}{0.778m}(m_a/10^{-5}$eV).
}
\label{tab:charge_on_plate}
\end{table}

\section{Sensitivity Prospects.}

\subsection{Narrow-band scheme}

As we discussed previously, a cavity has its optimal working frequency for signal pickup. A narrow band measurement scheme involves frequency filtering at this favored frequency range, and there are two popular ways to achieve this: (1) a high geometric quality factor of the cavity $Q_c$~\cite{Sikivie:1983ip}; (2) electronic filtering with an LCR circuit~\cite{Sikivie:2013laa} with a quality factor $Q_{c}= (\omega C R_s)^{-1}$, where $C, R_s$ denote the LCR's capacitance and resistance at the resonant point. In either case, when the signal's and the pickup's quality factors match each other, the high pickup quality factor is capable of enhancing the output power from a monochromatic perturbation source by a factor of the order ${\cal O}(Q_c)$. Assuming this condition could be met, the enhanced signal current is 
\be 
I_{a} =  Q_{c}\cdot q_0 \omega \cos(\omega t),
\ee
where $q_0$ is maximal (un-enhanced) charge build-up on the plates due to the signal field, and it is derived by a surface integral of {\bf E} field strength around one-quarter of the cylindrical plate's inner surface:
\ckk{
\be 
q_0 =\int_\text{surface} \epsilon_0 \vec{E}\cdot {\rm d}{\vec {A}},
\ee
%\ckk{where $1/2$ factor from $\phi$-alignment is canceled by having two sets of quarter-plates.} 
For the quadrupole configuration, this charge can be rewritten 
\be
\begin{array}{rcl}
    q_0& =  \int \epsilon_0E_r (\omega) \cdot dA  \simeq  \epsilon_0  E_r(r=R_0) \pi RL\\
%   &    =CU \simeq \eta_J\cdot \left.\frac{ \pi R L}{2}E_r(\omega)\right|_{r=R_{\rm cav},\phi=45^{\circ}} \\
  %   &\sim & B_0 h_0 c (\wg R_{\rm cav})^2, \\
\end{array}
\label{eq:q0} 
 \ee
Now we can make use of the fact that quarter-plates have an effective capacitance $C \sim \pi L \eta_J(\omega)$, and $\eta_J$ is a form factor to account for the geometric layout of the quadruple-shaped cavity. %Around the base mode $\omega\sim \omega_{211}$ this form factor is ${\cal O}(1)$ and it will decrease quickly if $\omega$ is away from $\omega_{211}$. 
From EM simulation results in Table~\ref{tab:charge_on_plate}, we can evaluate $\eta_J$ to be $1.10\epsilon_0 =1.10\times 8.85 \text{pF/m}$ for a 1 cm$^3$ cavity with $L=5cm$ and and $1.12\epsilon_0 (\text{pF/m})$ for a 1 m$^3$ cavity, respectively. Notice $\epsilon_0$ is $8.85$pf/m, having the same unit as $\eta_J$. We can estimate by $E_r(r=R_0) \sim B_0 h_{GW} (\wg R_0)^2$ at the center of each plate from dimension analysis. For a GHz-frequency GW wave with a stress intensity $h\sim 10^{-20}$, we will have $E_r\sim 10^{-10}$~V/m for a meter scale $R_{\rm cav}=1$ m cavity for a rough dimension analysis. Notice here we also assume we are at the resonance frequency with $\omega R \sim 1$. As we have been arguing below, we only work at most 1 or 2 orders of magnitude below the resonance frequency to get a reasonable sensitivity. Due to the imperfect symmetry of our cavity, we resort to numerical simulation to evaluate the signal electric fields. 
%The enhanced signal power is 
%\be
%P_{\rm sig.}= Q_c\cdot \vev{q^2}\omega~C^{-1}, \ckk{\text{how comes this?}}
%\ee
%where $C$ is the effective capacitance of the plates. 
For an LCR-type of filtering as shown in Fig.~\ref{fig:quarterly}, The $Q$-enhanced signal power at the resonance point is 
\be 
P_{\rm sig.}= \vev{ I_a^2}\cdot R_s =\frac{(Q_c \omega q_0)^2}{2Q_c\omega C}=Q_c\cdot \vev{q_0^2}\omega~(2C)^{-1},
\label{plott}
\ee
where $C$ is the effective capacitance of the plates, and this power equals the maximal thermal dissipation power at the LCR circuit. With the GW estimation, we have
\be
P_{\rm sig. }\sim Q_c \wg \frac{(  \epsilon_0 \pi R L B_0 h_0 )^2}{2 \pi L \eta_J} = Q_c \wg \pi R^2 L \frac{(  \epsilon_0 B_0 h_0   )^2}{2 \eta_J}.
\ee
Here the maximal signal power is for $\omega_g \sim 0.1 \omega_{211}$. Notice here is just for the estimation. We have used Eq.\ref{plott} and value from Table \ref{tab:charge_on_plate}  to plot the sensitivity curve.
}

\begin{figure}[t]
    \centering
    \includegraphics[scale=0.5]{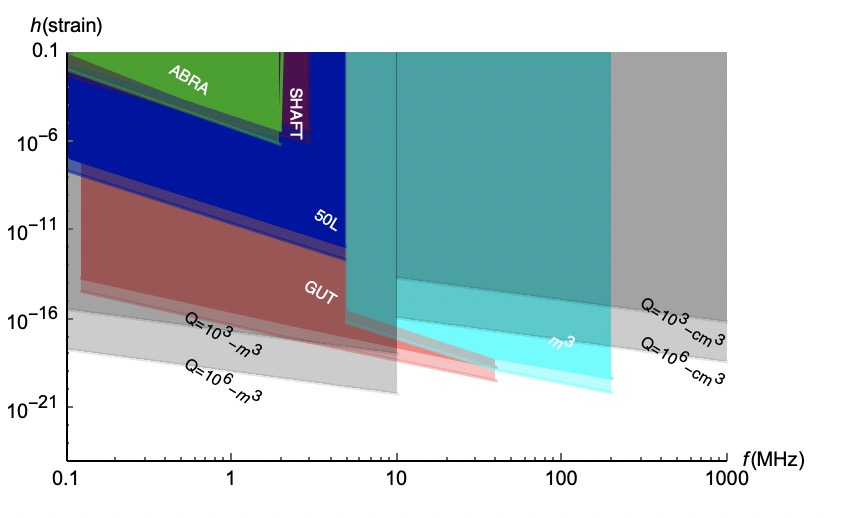}
    \caption{\ck{Projected sensitivity to the stress of a GW signal,  with coherent factor Q=$10^6$ and $10^3$, corresponding to narrowband searches. For comparison, existing experimental limits from ABRACADABRA\cite{ABRACADABRA:2018rtf,Salemi:2019xgl}, SHAFT\cite{Gramolin:2020ict}, and the projected sensitivity from DM-Radio\cite{Brouwer:2022bwo,Silva-Feaver:2016qhh,SnowmassOuellet,SnowmassChaudhuri} are also shown in plot.}
    }
    \label{fig:results}
\end{figure}

The signal power can be subsequently amplified and read out by cryogenic detectors. For a conceptual discussion in this work, we will not go into depth into noise discussion and only assume a thermal-noise-dominated background. At sub-GeV frequencies, the quantum noise is subdominant to amplifier noises, and a reasonable choice with modern cryogenic technique is an equivalent noise temperature at $T_N=0.1$ K~\cite{Duan:2022nuy}. This noise temperature gives the noise power of $P_{N}=k_B T_N\Delta f$, where $k_B$ is the Boltzmann constant, and for narrow-band filtering, we choose the bandwidth to be $\Delta f\sim \omega/(2\pi Q_c)$. The signal-to-noise rate is then
 \be
 {\rm SNR} = \frac{P_{\rm sig.}}{k_{\rm B}T_{N}}\sqrt{\frac{\Delta{t}}{\Delta f}},
 \ee
where $\Delta t$ is the observation time which we take to be 1 min here and we can require SNR$=3$ for a $3\sigma$ sensitivity criterion, the corresponding sensitivity on the GW stress is plotted in Fig. \ref{fig:results}.
%\begin{figure}[t]
%    \centering
%    \includegraphics[scale=0.4]{deltaphi.pdf}
%    \caption{The phase difference between the charge and two adjacent plates voltage difference. }
%    \label{fig:results_phase}
%\end{figure}

Note in our narrow-band scheme there is a {\it not-small} assumption that the signal should be also narrow-band in its frequency domain, to match that of the pickup and trigger the high-$Q$ enhancement. This requires a near monochromatic signal and a relatively long coherence time scale, at least longer than $2\pi Q_c/\omega$ to ring up the resonance enhancement. Typically, this indicates the narrow-band measurement scheme is best for a GW source with good coherent oscillation periodicity. There are potential candidates of such sources, for instance, the GW induced by the coherent oscillation of a light scalar field as a dark matter condensate~\cite{Sun:2020gem} or a superradiance cloud around Kerr black holes~\cite{Brito:2015oca}, etc. For non-periodic yet coherent sources, such as transient events, we will change our plan and consider the broad-band $\Delta f \sim f$ sensitivity of the non-resonant measurement.

%From the dimension analysis, $E_r$ with quadrupole solution, 

\subsection{Non-resonant scheme}

\ckk{When the signal spectrum has a broader frequency spread, the signal rate will decrease at a sharply-turned pickup. In the case of GW signals, there is also interest in case the spectral shape is not highly monochromatic, and one can turn up the resistance in the LCR filter accordingly to match the signal's reduced Q value. That would be the case for non-periodic transient signals. The sensitivity of our measurement can be estimated by comparing the signal-to-noise ratio within the signal's expected time duration. When the signal has an extended spectrum in the frequency domain, the choice of measurement bandwidth needs to account for the frequency dependence of the quarterly split pickup.}

GW-induced effective current $j_{\rm eff}$ is generated by time-variance in the metric $\partial_t h$, the induced current between electrodes is proportional to $\partial_t q\propto \partial_t j_{\rm eff}$, therefore our measured signal responds to $\partial_t^2 h$ and should have a higher sensitivity to the high-frequency part of the GW signal spectrum. For specific sources, the shape of $\partial_t^2 h$ will affect our detection efficiency: we need to choose our bandwidth to cover the frequency range where $\partial_t^2 h$ maximizes. 

Highly motivated high-frequency coherent GW sources include the binary mergers of primordial black holes and blackhole superradiance. For the blackhole binary merger, the emitted gravitational wave frequency is associated with the ISCO and the inspiral emission peaks at
\be
f _{\rm ISCO}\simeq 4.4\text{MHz} \, \left(\frac{10^{-3} M_\odot}{m^\PBH_1+m^\PBH_2} \right) \!.
\label{eq:fISCO}
\ee
and we have the GW strains from the merger along the symmetry axis of the circular orbit 
~\cite{maggiore}
\be
h_{+,\times}^\PBH \sim  2.5\times 10^{-21} \left( \frac{10~\text{kpc}}{D}\right)  \left(\frac{M_\PBH}{10^{-3} M_\odot}\right)^{5/3} \left( \frac{ f}{1.9~\text{MHz}} \right)^{2/3}\!.
\label{eq:hBinary}
\ee
where $M_{PBH}=(m_1m_2)^{3/5}/(m_1+m_2)^{1/5}$ is the binary chirp mass. \ck{The GW signal from an individual merger event is not monochromatic; its spectrum may spread over a few orders of magnitude in the frequency domain.
For good detection efficiency, we assume the cylinder's $\omega_{211}$ roughly matches a target GW signal's characteristic frequency and consider a wide frequency neighborhood. For this purpose, we tune down the LC quality factor to $Q_c \sim1$. 
The cavity's frequency response, or $\xi(\omega)\equiv q(\omega)/q(\omega_{211})$ for a fixed $h_0$, is expected to drop significantly when $\omega \ll \omega_{211}$. In the $\omega >\omega_{211}$ direction, signal decoherence will suppress the geometric factor if the cavity dimension is multiples of the optimal mode-211 wavelength. Therefore, a reasonable choice of the cavity's frequency width is $\Delta \omega \sim \omega$ near the TE$_{211}$ mode. Therefore, if we assume a very broad GW spectrum of $h(\omega)$, we can make the approximation that $I_a(\omega)\approx \xi(\omega)\cdot I_a(\omega_{211})\cdot$ inside the $\Delta \omega =\omega$ window and ignore any $I_a(\omega)$ outside this frequency range.}
%The cavity's frequency response, or $\xi(\omega)\equiv q(\omega)/q(\omega_{211})$ for a fixed $h_0$, is expected to drop significantly when $\omega \ll \omega_{211}$. In the $\omega >\omega_{211}$ direction, signal decoherence will suppress the geometric factor if the cavity dimension is multiples of the optimal mode-211 wavelength. Therefore, a reasonable choice of the cavity's frequency width is $\Delta \omega \sim \omega$ near the TE$_{211}$ mode. Therefore, if we assume a very broad GW spectrum of $h(\omega)$, we can make the approximation that $I_a(\omega)\approx \xi(\omega)\cdot I_a(\omega_{211})\cdot$ inside the $\Delta \omega =\omega$ window and ignore any $I_a(\omega)$ outside this frequency range.}

%As  $\eta_J(\omega)$ is expected to decrease rather quickly when $\omega$ deviates away from $\omega_{211}$, the geometry factor would effectively act as a filter and determines the effective bandwidth when $\partial_t^2 h$ is more extensive than $\eta_J(\omega)$. \ckk{From EM simulations, we have $\Delta \omega = {\cal O}(1) \omega_{211}$,} beyond which $I_a$ will become suppressed by more than one order of magnitude. 

\ck{Here we choose $\Delta\omega=\omega$ to offer a proof-of-principle estimate. For specific sources, if $\partial_t^2 h$ shows a narrower peak than the cavity's bandwidth, one would then tune up the LC quality factor that suits the particular spectrum shape.  In practice, $\Delta \omega$ will also be limited by the maximal linear amplification bandwidth of the amplifier/readout system.}

 %\ckk{$\eta_J(\omega)$ is constant through $\omega$, see table I.}

\begin{figure}[h]
    \centering
 \includegraphics[width=13cm]{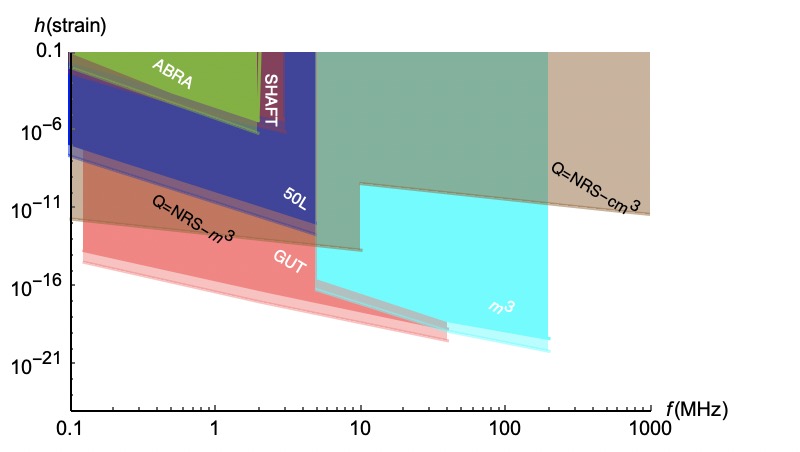}
    \caption{SNR=3 sensitivity for transient events, like PBH merger, and BH superradiance collapse. We use a bandwidth of $\Delta f =f$ for sensitivity estimate.}
    \label{fig:results2}
\end{figure}

For occasional signals, the SNR can be estimated by comparing the signal and noise power during the characteristic time-duration $\Delta t$ of the signal, 
\be
{\rm SNR}=\frac{\left<P_{\rm sig.}\right>}{P_{\rm noise}} = \frac{
R\cdot \sum\limits_{n=-\infty }^{+\infty}|c_n|^2 
}
{k_B T_N \Delta f}, 
\label{eq:snr_discrete_ft}
\ee
where the numerator is the time-averaged signal power and it makes use of Parseval's Theorem, i.e. $\int |I_a(t)|^2{\rm d}t =\Delta t \cdot \sum_n |c_n|^2$. $R$ is the pickup circuit's resistance and $R\sim(\omega C)^{-1}$ for $Q\sim 1$. $c_n$ is the complex Fourier transform coefficient over the finite $\Delta t$ time window,
\be 
c_n \equiv \frac{1}{\Delta t}\int_0^{\Delta t} I_a(t) e^{i\frac{2\pi}{\Delta t}nt}{\rm d}t, ~~~\ n =1,2,...  
\ee
Note that $h(\omega)$ in Eq.~\ref{eq:hBinary} is often given by a continuous transformation~\cite{maggiore} using an unnormalized basis $h(\omega)\equiv \int\limits^{^+\infty}_{-\infty}h(t)e^{i\omega t}{\rm d}t$. Technically, if one assumes a periodic condition for normalization, or $h(t)=h(t+\Delta t)$, the summation of $c_n(\omega)$ over discrete $\omega_n$ can be replaced by an integral in the $\Delta t\rightarrow +\infty$ limit: $\frac{2\pi}{\Delta t}\sum |I_a(\omega_n)|^2\rightarrow \int |I_a(\omega)|^2{\rm d}\omega$, so that Eq.~\ref{eq:snr_discrete_ft} can be rewritten as
\be 
\frac{\left<P_{\rm sig.}\right>}{P_{\rm noise}} = \frac{R}{2\pi \Delta t}\frac{2\int_{\Delta \omega}  |\tilde{I}_a(\omega)|^2~{\rm d}\omega}
{k_B T_N \Delta f}.
\ee
The factor of 2 comes from negative frequency. Note $\Delta t$ in the denominator normalizes $I_a(\omega)$, thus it does not indicate for higher sensitivity over an arbitrarily short time window. For mergers, we take $\Delta t=1$ s for a typical merger signal duration. 
For noises, we also assume an effective noise temperature  $T_N =0.1$~K. The expected SNR=3 sensitivity for the cm$^3$ and m$^3$ cavity dimensions are illustrated in Fig.~\ref{fig:results2}.

We follow the same treatment for the GW signals from blackhole's axion superradiance collapse~\cite{Brito:2015oca}. The axion can form clouds around the black holes when the axion Compton wavelength is comparable to the Schwarzschild radius. Then either the axion decay\cite{Sun:2020gem} or axion annihilation can emit the gravitational waves with a frequency set by the axion mass scale:%
\be
\frac{f}{1~\text{MHz}} \sim  \frac{m_a}{10^{-5}~\text{eV}} \sim  \frac{10^{-5} M_\odot}{m_\text{BH}}.
\ee
Then we can roughly estimate the GW amplitude
\be
h \sim 10^{-24}\, \frac{1~\text{MHz}}{f} \frac{10~\text{kpc}}{D},
\ee
Unlike the single-frequency scalar oscillation, the superradiance collapse is a quick and sudden event that ends on a short time scale $t_{\rm col.}$ no more than a few axion oscillation periods. In principle, this causes a frequency spread $\delta f\sim t_{\rm col.}^{-1}/2\pi$ due to the lack of a long coherent time duration. Thus we also assume a $\Delta f\sim f$ bandwidth for such events, and the corresponding sensitivity is shown as in Fig.~\ref{fig:results2}.

Mergers and superradiance collapses are two examples of potential nearly coherent sources, with short durations and certain directionality, which are viable sources for our detection scheme. We may also estimate the event rate around different sky angles, which we leave for future work. 

Here we also comment on the sensitivity of the stochastic sources, and refer to Ref.~\cite{Aggarwal:2020olq} for detailed discussions. Many beyond the Standard Model scenarios in the early Universe can source stochastic gravitational waves, e.g. phase transition, preheating or reheating, and topological defects dynamics. Since gravitational waves play the same role as early universe radiation in the cosmic evolution, they are bounded by the effective number of additional neutrinos $\Delta N_\text{eff}$ as $h < 10^{-27} (1\text{MHz}/f)\Delta N_\text{eff}^{1/2} $ \cite{Pisanti:2020efz,Yeh:2020mgl}. This means the amplitude of the stochastic gravitational wave is a few orders smaller than the coherent ones in general. 

Especially, the signal from stochastic sources with random phases resembles white noise in one detector. One solution to identify such stochastic signals is through multiple detectors, in light of the pulsar timing array sensitivity of the stochastic gravitational waves. Considering our measurement scheme is electric in nature, two detectors moderately separated with cross-correlation of the electric signals may provide a certain signal-to-noise ratio of the stochastic background.
%\ckk{\bf [not sure what you mean by enough?]}
%These examples illustrate that different search strategies will need to be implemented to optimally search for different possible GW sources.
%
%Generally, however, the GW signal is expected to be less coherent than axion dark-matter, and so in general strategies searching for broader signals in the frequency domain will need to be devised.
%
%(In this sense there are strong similarities with the search for relativistic axions, for further discussion see Ref.~\cite{Dror:2021nyr}.)

\section{Conclusion}

\ck{To summarize, we consider a cryogenic detection scheme of the electric signals induced by gravitational waves via the inverse Gertsenshtein effect~\cite{Gertsenshtein1962}, inside a strong static magnetic field. We make use of the transverse electric mode inside a quarterly-split cylinder geometry to maximize the signal induction from the oscillatory effective currents from gravitational waves. The current signal derives from the electric charge build-up on the surface of the cavity plates. We have performed EM simulation with the quarterly-split geometry and obtained the cavity's form factor below the frequency of TE$_{211}$ mode, which characterizes the frequency response of our experimental setup. In the case of a narrow-frequency gravitational wave, the signal can be further enhanced by a large quality factor $Q$ that matches the bandwidth of the gravitational wave, and the corresponding strain sensitivity is projected to around $h\sim 10^{-20}\cdot (10^6/Q)$ for a meter-sized split cavity. For signals more extended in the frequency domain, we consider a broad-band signal with a characteristic frequency range wider than the cavity's geometric bandwidth and make an estimate of our design's sensitivity in broadband, non-resonant ($\Delta f \sim f$) mode. The broadband sensitivity is found to be $h\sim 10^{-13}$ for a meter-scale cavity.}
\bigskip
\bigskip
%\vspace{-10pt}
\begin{acknowledgments} \vspace{-10pt}
The authors thank for the support from the National Natural Science Foundation of China (Nos. 12105013, 12150010) and
International Partnership Program of the Chinese Academy of Sciences for Grand Challenges (112311KYSB20210012).
%Y.G. is supported under grant No. 12150010 supported by the National Natural Science Foundation of China. \ckk{Your grants}

%\bigskip
%\bigskip
\iffalse
{\bf{Conflict of interest:}} The authors declare that they have no conflict of interest.
{\bf{Author contribution:}} Junxi Duan and Yu Gao work out the experimental details. Chang-Yin Ji performs the numerical simulation. Yu Gao, Sichun Sun, Yugui Yao and Yun-Long Zhang initiate the project. All authors participate in the draft editing.
\fi
%\begin{figure}[h]
%    %\centering
%    \includegraphics[scale=0.4]{JiCY1.jpg}
%    \caption{    } \label{fig1}
%\end{figure}
\end{acknowledgments}

\bibliography{refs}

\end{document}